\documentclass[aps,prl,twocolumn,superscriptaddress,preprintnumbers,showpacs]{revtex4}
\usepackage{graphics}
\def\be{\begin{equation}}
\def\ee{\end{equation}}
\def\bea{\begin{eqnarray}}
\def\eea{\end{eqnarray}}
\def\car{{\cal C}}
\def\Tr{{\rm Tr\,}}
\def\car{{\cal R}}
\def\LN{\Lambda_{\rm NC}}

\begin{document}


\title{Improved $Z\,\rightarrow \,\gamma  \gamma$ decay in the
renormalizable gauge sector of \\the noncommutative standard model}

\author{M. Buri\' c}
\affiliation{Faculty of Physics, University of Belgrade,
P. O. Box 368, 11001 Belgrade, Serbia}
\author{D. Latas}
\affiliation{Faculty of Physics, University of Belgrade,
P. O. Box 368, 11001 Belgrade, Serbia}
\author{V. Radovanovi\' c}
\affiliation{Faculty of Physics, University of Belgrade,
P. O. Box 368, 11001 Belgrade, Serbia}
\author{J.~Trampeti\' c}
\affiliation{Theoretical Physics Division,
Rudjer Bo\v{s}kovi\'{c} Institute, P.O.Box 180,
10002 Zagreb, Croatia}
\date{\today}

\begin{abstract}
In this work we propose the $Z\rightarrow \gamma\gamma$
decay as a process strictly
forbidden in the standard model,
suitable for the search of  noncommutativity
of coordinates at very short distances.
We computed the $Z\rightarrow \gamma\gamma$
partial width in the framework of the recently proposed one-loop
renormalizable gauge sector of the noncommutative standard model.
New experimental possibilities at LHC are analyzed and a
firm bound to the scale of noncommutativity parameter is set around 1 TeV.
\end{abstract}


\maketitle


In this work we consider the standard-model (SM) forbidden $Z \to \gamma\gamma$ decay
originating from the renormalizable
gauge sector of the noncommutative standard model (NCSM) given in
\cite{Buric:2006wm},
which could be probed in high-energy collider experiments.

The possibility of the noncommutative structure
of space-time is of interest in its own right
and its experimental discovery would be a result of fundamental importance.
The simplest way to introduce  noncommutativity (NC)
is to replace the usual  coordinates $x^\mu$ by the  noncommutative coordinates
$\hat x^\mu$ which obey the commutation rule
\begin{equation}
\left[{\hat x}^{\mu},{\hat x}^{\nu} \right]=i h \theta^{\mu\nu}, \quad
\left[\theta^{\mu\nu},{\hat x}^{\rho} \right]=0.
\label{CR}
\end{equation}
The matrix $\theta^{\mu\nu}$ is  constant, real, and antisymmetric.
The algebra defined by (\ref{CR})  is  called noncommutative Minkowski space.
The fields $\hat\Phi$ on this space are functions of noncommutative coordinates,
$\hat\Phi (\hat x^\mu)$; they can be represented by functions of
ordinary commuting coordinates if the multiplication
is represented by the Moyal-Weyl $\star$-product.
The constants $\theta^{\mu\nu}$ in (\ref{CR}) are dimensionless: the
noncommutativity scale $\Lambda_{\rm NC}$  defined by
$ h =\Lambda_{\rm NC}^{-2}$ has the dimension
of the inverse length or energy. In principle, the order of magnitude
of $\LN^{-1}$ can be between the proton radius and the Planck length.

Our motivation to reexamine the partial width of the
$Z\,\rightarrow \,\gamma  \gamma$ decay  and,  in general,
to reexamine the construction of
gauge-field models on noncommutative Minkowski space
\cite{Seiberg:1999vs,Madore:2000en,Maja,Bichl:2001cq}, was a recent
result on the one-loop renormalizability of
the $\theta$-expanded  $\rm SU(N)$ gauge theory obtained in
\cite{Buric:2005xe,Latas:2007eu}.
In \cite{Buric:2006wm} we analyzed the SM and
succeeded in constructing a model which had the renormalizable
gauge sector to $\theta$-linear order.
These interactions proved to be free of divergences \cite{Buric:2006wm}
and anomaly free whenever its commutative counterpart
is anomaly free \cite{Brandt:2003fx}.

Experimental evidence for noncommutativity coming
from the gauge sector should be searched for
in processes which involve these interactions.
The simplest and most natural choice is the
$Z\rightarrow \gamma\gamma$ decay, kinematically
allowed for on-shell particles. As it is
forbidden in the SM by angular-momentum conservation and Bose statistics
(Landau-Pomeranchuk-Yang Theorem) \cite{Yang}, it
would serve as a clear signal for the existence of
space-time noncommutativity. However, note  that NC interaction, for
 fixed $\theta^{\mu\nu}$,  breaks  Poincar\' e symmetry.
Signatures of noncommutativity were
discussed previously within particle physics in
\cite{Hewett:2000zp,Peter,Abbiendi:2003wv,Ohl:2004tn,Carroll:2001ws,Mocioiu:2000ip,Josip}.

The action of NC gauge theory analogous to
that of the ordinary Yang-Mills theory is given in
\cite{Calmet:2001na,Aschieri:2002mc,Behr:2002wx,Duplancic:2003hg,Melic:2005fm},
with the commutative field strengths replaced by the noncommutative ones.
However, in this work we start with
a higher order noncommutative gauge field theory \cite{Latas:2007eu},
\begin{equation}
 S =- \frac{1}{2}{\rm Tr}\int d^4x
\left(1-\frac{a-1}{2}\,h\, \theta^{\mu\nu}\,
\widehat F_{\mu\nu}\right)\star\widehat F_{\rho\sigma}
\star\widehat F^{\rho\sigma}.
\label{action1}
\end{equation}
This action, after expansion in commutative fields
(Seiberg-Witten map), given to linear order  by
\begin{equation}
{\widehat F}_{\mu\nu}=F_{\mu\nu} +
\frac{h}{4}\,\theta^{\sigma\rho} \left(2\{F_{\mu\sigma}, F_{\nu\rho}\}
-\{V_{\sigma} ,(\partial_{\rho}+{\cal D}_{\rho})F_{\mu\nu} \} \right),
\label{fields1}
\end{equation}
leads to
\begin{eqnarray}
S &=&\Tr\int \mathrm{d}^4x\,\left(-\frac{1}{2}
F_{\mu\nu}F^{\mu\nu}
\right.
\nonumber\\
&+&
\left.
h\,\theta^{\mu\nu}\, (
\frac{a}{4}\, F_{\mu\nu}F_{\rho\sigma}
-F_{\mu\rho}F_{\nu\sigma} )F^{\rho\sigma}\right) .
\label{act}
\end{eqnarray}
Linear-order action (\ref{act}) is one-loop renormalizable only
for $a=1,3$ and for specific representations of the gauge
potential,
\cite{Buric:2006wm,Buric:2005xe,Latas:2007eu,Martin:2006gw}. The
initial action (\ref{action1}) is invariant under the NC gauge
transformations, while (\ref{act}) is manifestly invariant under
the ordinary gauge transformations. The  dependence of actions
(\ref{action1}) and (\ref{act}) on the {\it freedom parameter}
$a$, as can easily be seen, originates from the higher order gauge
interaction term $\,\theta^{\mu\nu}\widehat
F_{\mu\nu}\star\widehat F_{\rho\sigma}\star\widehat
F^{\rho\sigma}$.

As we have already mentioned, in \cite{Buric:2006wm}
renormalizability/finiteness points out the value $a=3$ as physical;
however,  we shall keep the value of $a$ arbitrary in calculations
and use $a=3$ at the end.

The trace in (\ref{action1}) and  (\ref{act}) is, in principle,
taken  over all irreducible representations with arbitrary weights.
Obviously, gauge models are representation dependent in the NC case:
the choice of  representations
has a strong influence on the theory, on both the form of interactions and
the renormalizability properties.

In order to obtain the triple neutral gauge boson interactions,
starting with the SM gauge potential
\cite{Calmet:2001na,Aschieri:2002mc,Behr:2002wx,Duplancic:2003hg,Melic:2005fm},
\begin{eqnarray}
V_{\mu}=g'{\cal A}_{\mu}Y + g\sum^3_{a=i}B_{\mu,i}T^i_L+g_s\sum^8_{b=1}G_{\mu,b}T^b_S ,
\label{V}
\end{eqnarray}
we choose to sum over all particle representations of the standard model,
i.e. we take nonzero weights in (\ref{act}) only for the SM particle representations.
There are five multiplets
of fermions and one Higgs multiplet for each generation.
However, owing to  symmetry arguments
(vanishing of the symmetric coefficients $d^{ijk}$ for SU(2) and
invariance of the color sector under the charge conjugation),
we obtain the action different from that found
preoviusly \cite{Behr:2002wx}.
This action to linear order in $\theta$ is \cite{Buric:2006wm}
\begin{eqnarray}
\lefteqn{S_{\rm gauge}= S_{\rm gauge}^{\rm SM}}
\label{action3} \\
& &\hspace{-7mm}{}+{g'}^3\kappa_1{\theta^{\rho\tau}}\hspace{-2mm}\int \hspace{-1mm}d^4x\,
\left(\frac{a}{4}f_{\rho\tau}f_{\mu\nu}-f_{\mu\rho}f_{\nu\tau}\right)f^{\mu\nu}
 \nonumber \\
& &\hspace{-7mm}{}+g'g^2\kappa_2 \, \theta^{\rho\tau}\hspace{-2mm}\int
\hspace{-1mm} d^4x \sum_{i=1}^{3}
\Big[(\frac{a}{4}f_{\rho\tau}B^i_{\mu\nu}-
2f_{\mu\rho}B^i_{\nu\tau})B^{\mu\nu,i}\!
\nonumber \\
& &\hspace{2.5cm}{}
+f^{\mu\nu}(\frac{a}{2}B^i_{\rho\tau}B^{i}_{\mu\nu}-B^i_{\mu\rho}B^{i}_{\nu\tau})
\Big]
 \nonumber \\
& &\hspace{-7mm}{}+g'g^2_s\kappa_3\, \theta^{\rho\tau}\hspace{-2mm}\int
\hspace{-1mm} d^4x \sum_{b=1}^{8}
\Big[(\frac{a}{4}f_{\rho\tau}G^b_{\mu\nu}-
2f_{\mu\rho}G^b_{\nu\tau})G^{\mu\nu,b}\!
\nonumber \\
& &\hspace{2.5cm}{}
+f^{\mu\nu}(\frac{a}{2}G^b_{\rho\tau}G^{b}_{\mu\nu}-G^b_{\mu\rho}G^{b}_{\nu\tau})
\Big] .
\nonumber
\end{eqnarray}
Here, $f_{\mu\nu}$, $B^i_{\mu\nu}$, and $G^b_{\mu\nu}$
are the physical field strengths which correspond
to  $\rm U(1)_Y$, $\rm SU(2)_L$, and $\rm SU(3)_C$, respectively.
The constant $a$ can be arbitrary and in the previous work it was fixed to 1.
The couplings $\kappa_1$, $\kappa_2$, and $\kappa_3$,
as functions of the weights $C_\car$, that is of the $C_i(=1/g_i^2),\; i=1,...,6$,
are parameters of the model.
Further details of the model and of constraints
imposed on $\kappa_1,\kappa_2$, $\kappa_3$,
and $C_i$ are given in \cite{Duplancic:2003hg}
and is not necessary to repeat them here.

The divergent one-loop vertex correction to (\ref{action3})
as a function of the freedom parameter $a$ is \cite{Buric:2006wm}
\begin{eqnarray}
\Gamma_{\rm div}&=& \frac {11}{3 (4\pi)^2\epsilon}\int d^4 x
\Big(B_{\mu\nu}^iB^{\mu\nu i} + \frac{3}{2} G_{\mu\nu}^a G^{\mu\nu a} \Big)
\nonumber \\
&+&
\frac{4}{3(4\pi)^2\epsilon}{g^\prime} g^2  \kappa_2 (3-a)
\nonumber \\
&\times&
\theta^{\mu\nu}\int d^4 x
\big(\frac{1}{4} f_{\mu\nu}B_{\rho\sigma}^i
- f_{\mu\rho}B_{\nu\sigma }^i \big) B^{\rho\sigma i}
\nonumber \\
&+& \frac{6}{3(4\pi)^2\epsilon}{g^\prime} g^2_S \kappa_3 (3-a)
\nonumber \\
&\times&
\theta^{\mu\nu}\int d^4 x
\big(\frac{1}{4} f_{\mu\nu}G_{\rho\sigma }^a
- f_{\mu\rho}G_{\nu\sigma }^a \big)G^{\rho\sigma a}\,.
\label{div}
\end{eqnarray}
From the result (\ref{div}) it is clear that the expanded gauge action
(\ref{action3}) is renormalizable
only for the value $a=3$. Even more,
its noncommutative part is finite or free of divergences, so
the noncommutativity parameter $\theta$ need not be renormalized.
A detailed description of the renormalization procedure,
including the results for the bare fields and couplings, is given in
\cite{Buric:2006wm}.

Finally, from the action (\ref{action3}) we extract the
triple-gauge boson terms which are not present in the commutative SM Lagrangian.
In terms of the physical fields $A,\ W^{\pm},\ Z$, and $ G$ they are
\begin{eqnarray}
{\cal L}^{\theta}_{\gamma\gamma\gamma}&=&\frac{e}{4}\sin2{\theta_W}\;{\rm K}_{\gamma\gamma\gamma}
{\theta^{\rho\tau}}
\nonumber\\
& & \times A^{\mu\nu}\left(aA_{\mu\nu}A_{\rho\tau}-4A_{\mu\rho}A_{\nu\tau}\right),
\label{L1}\\
{\rm K}_{\gamma\gamma\gamma}&=&\frac{1}{2}\; gg'(\kappa_1 + 3 \kappa_2);
\nonumber\\
& & \nonumber \\
{\cal L}^{\theta}_{Z\gamma\gamma}&=&\frac{e}{4} \sin2{\theta_W}\,{\rm K}_{Z\gamma \gamma}\,
{\theta^{\rho\tau}}
\nonumber\\
& & \times \left[2Z^{\mu\nu}\left(2A_{\mu\rho}A_{\nu\tau}-aA_{\mu\nu}A_{\rho\tau}\right)\right.
\nonumber\\
& & +\left. 8 Z_{\mu\rho}A^{\mu\nu}A_{\nu\tau} -aZ_{\rho\tau}A_{\mu\nu}A^{\mu\nu}\right],
 \label{L2}\\
{\rm K}_{Z\gamma\gamma}&=&\frac{1}{2}\; \left[{g'}^2\kappa_1+\left({g'}^2-2g^2\right)\kappa_2\right];
\nonumber\\
& &\nonumber \\
{\cal L}^{\theta}_{ZZ\gamma}&=&{\cal L}_{Z\gamma\gamma}(A\leftrightarrow Z),
\label{L3}\\
{\rm K}_{ZZ\gamma}&=&\frac{-1}{2gg'}\;\left[{g'}^4\kappa_1+g^2\left(g^2-2{g'}^2\right)\kappa_2\right];
\nonumber \\
& &\nonumber \\
{\cal L}^{\theta}_{WW\gamma}&=&\frac{e}{4} \sin2{\theta_W}\,
{\rm K}_{WW\gamma}\,{\theta^{\rho\tau}}
\nonumber\\
& & \times\left\{A^{\mu\nu}\left[2\left(W^+_{\mu\rho}W^-_{\nu\tau}
+W^-_{\mu\rho}W^+_{\nu\tau}\right)\right.\right.
\nonumber\\
& & -\left.\left. a\left(W^+_{\mu\nu}W^-_{\rho\tau}
+W^-_{\mu\nu}W^+_{\rho\tau}\right)\right]\right.
\nonumber\\
& & +\left.\left.4 A_{\mu\rho}\left(W^{+\mu\nu}W^-_{\nu\tau}+W^{-\mu\nu}W^+_{\nu\tau}\right)
\right.\right.
\nonumber\\
& & \left.\left.- aA_{\rho\tau}W^+_{\mu\nu}W^{-\mu\nu}\right]\right\}  ,
\label{L4}\\
{\rm K}_{WW\gamma}&=&-\frac{g}{g'}\; \left[{g'}^2+g^2\right]\kappa_2 \,,
\nonumber
\end{eqnarray}
where $A_{\mu\nu} \equiv \partial_{\mu}A_{\nu}-\partial_{\nu}A_{\mu}$, etc.
The structure of the other interactions as  $WWZ$, $ZZZ$, $Zgg$, and $\gamma gg$ is given in
\cite{Behr:2002wx,Duplancic:2003hg,Melic:2005fm}.

In this letter we focus on the branching ratio of the
$Z\rightarrow \gamma\gamma$ decay in renormalizable model (\ref{L2}).
Note that each term from the
$\theta$-expanded action (\ref{action3}) is manifestly invariant
under the ordinary gauge transformations.
The gauge-invariant amplitude
${\cal A}^{\theta}_{Z\rightarrow \gamma\gamma}$
for the $Z(k_1)\rightarrow\gamma(k_2)\,\gamma(k_3)$ decay
in the momentum space reads
\begin{eqnarray}
{\cal A}^{\theta}_{Z\to \gamma\gamma}=&-&2e \sin2{\theta_W}\,{\rm K}_{Z\gamma \gamma}
{\Theta^{\mu\nu\rho}_3}(a;k_1,-k_2,-k_3)
\nonumber\\
&\times&
 \epsilon_{\mu}(k_1) \epsilon_{\nu}(k_2) \epsilon_{\rho}(k_3).
\label{ampl}
\end{eqnarray}
The tensor ${\Theta^{\mu\nu\rho}_3}(a;k_1,k_2,k_3)$ is given by
\begin{eqnarray}
{\Theta^{\mu\nu\rho}_3}(a;k_1,k_2,k_3)&=&
-\,(k_1 \theta k_2)\,
\nonumber \\
& &
\hspace*{-3cm}
\times
[(k_1-k_2)^\rho g^{\mu \nu} +(k_2-k_3)^\mu g^{\nu \rho} + (k_3-k_1)^\nu g^{\rho \mu}]
\nonumber \\
& &
\hspace*{-3cm}
-\,\theta^{\mu \nu}\,
[ k_1^\rho \, (k_2 k_3) - k_2^\rho \, (k_1 k_3) ]
\nonumber \\
& &
\hspace*{-3cm}
-\,\theta^{\nu \rho}\,
[ k_2^\mu \, (k_3 k_1) - k_3^\mu \, (k_2 k_1) ]
\nonumber \\
& &
\hspace*{-3cm}
-\,\theta^{\rho \mu}\,
[ k_3^\nu \, (k_1 k_2) - k_1^\nu \, (k_3 k_2) ]
\nonumber \\
& & \hspace*{-3cm}
+\,(\theta k_2)^\mu \,\left[g^{\nu \rho}\, k_3^2 - k_3^\nu k_3^\rho\right]
+(\theta k_3)^\mu\,\left[g^{\nu \rho}\, k_2^2 - k_2^\nu k_2^\rho\right]
\nonumber \\
& & \hspace*{-3cm}
+\,(\theta k_3)^\nu \,\left[g^{\mu \rho}\, k_1^2 - k_1^\mu k_1^\rho \right]
+(\theta k_1)^\nu \,\left[g^{\mu \rho}\, k_3^2 - k_3^\mu k_3^\rho \right]
\nonumber \\
& & \hspace*{-3cm}
+\,(\theta k_1)^\rho \,\left[g^{\mu \nu}\, k_2^2 - k_2^\mu k_2^\nu \right]
+(\theta k_2)^\rho \,\left[g^{\mu \nu}\, k_1^2 - k_1^\mu k_1^\nu \right]
\nonumber \\
& & \hspace*{-3cm}
+\,\theta^{\mu\alpha}(ak_1+k_2+k_3)_{\alpha} \,\left[g^{\nu \rho}\,(k_3 k_2)-k_3^\nu k_2^\rho\right]
\nonumber \\
& & \hspace*{-3cm}
+\,\theta^{\nu\alpha} (k_1+ak_2+k_3)_{\alpha} \,\left[g^{\mu \rho}\,(k_3 k_1)-k_3^\mu k_1^\rho\right]
\nonumber \\
& & \hspace*{-3cm}
+\,\theta^{\rho\alpha} (k_1+k_2+ak_3)_{\alpha} \,\left[g^{\mu \nu}\,(k_2 k_1)-k_2^\mu k_1^\nu\right]
\, ,
\label{ampli}
\end{eqnarray}
where the 4-momenta $k_1,k_2,k_3$ are taken to be incoming, satisfying the momentum
conservation $(k_1+k_2+k_3=0)$.
In (\ref{ampli}) the freedom parameter $a$
appears symmetric in physical gauge bosons which enter the interaction point,
as one would expect.
For $a=1$, the tensor (\ref{ampli})
becomes  the tensor $\Theta_3((\mu,k_1),(\nu,k_2),(\rho,k_3))$ from \cite{Melic:2005fm}.

The amplitude (\ref{ampl})
with the Z boson at rest gives the total rate
for the $Z \rightarrow \gamma\gamma$ decay:
\begin{eqnarray}
&&\Gamma_{Z\rightarrow \gamma\gamma}
 =\frac{\alpha}{4} \frac{M^5_Z}{\Lambda^4_{\rm NC}} \sin^2 2\theta_W {\rm K}^2_{Z\gamma \gamma}
\label{eqn2}\\
&&\times
\frac{1}{18}\left[(13a^2-50a+51){\vec E}_{\theta}^2+(a^2+2a+3){\vec B}_{\theta}^2\right] ,
\nonumber
\end{eqnarray}
where ${\vec E}_{\theta}=\{{\theta}^{01},{\theta}^{02},{\theta}^{03}\}$ and
${\vec B}_{\theta}=\{{\theta}^{23},{\theta}^{31},{\theta}^{12}\}$
are dimensionless coefficients of order one, representing the time-space and space-space
noncommutativity, respectively. For the $Z$ boson at
rest, polarized in the direction of the third axis, we obtain the
following {\it polarized} partial width:
\begin{eqnarray}
\Gamma^3_{Z\rightarrow \gamma\gamma}
 &=& \frac{\alpha}{60} \frac{M^5_Z}{\Lambda^4_{\rm NC}}
\sin^2 2\theta_W {\rm K}^2_{Z\gamma \gamma}
\nonumber\\
&& \times \Big[\frac{1}{4}(7a^2-38a+55)\left((E_\theta^1)^2+
(E_\theta^2)^2\right) \nonumber\\ && \phantom{\times \Big[}
+(29a^2-106a+100)(E_\theta^3)^2
\nonumber\\
&&  \phantom{\times \Big[} +
\frac{1}{4}(a-1)^2\left((B_\theta^{1})^2+ (B_\theta^{2})^2\right)
\nonumber\\
&&  \phantom{\times \Big[}+(2a^2+6a+7)(B_\theta^{3})^2 \Big].
\label{eqn1}
\end{eqnarray}
For $a=3$, we have
\begin{eqnarray}
\Gamma_{Z\rightarrow \gamma\gamma}
&=&\frac{\alpha}{4} \;\frac{M^5_Z}{\Lambda^4_{\rm NC}}\; \sin^2 2\theta_W
{\rm K}^2_{Z\gamma \gamma}
\big({\vec E}_{\theta}^2 + {\vec B}_{\theta}^2\big),
\label{eqn3}
\end{eqnarray}
and
\begin{eqnarray}
\Gamma^3_{Z\rightarrow \gamma\gamma} &=&\frac{\alpha}{60}
\;\frac{M^5_Z}{\Lambda^4_{\rm NC}}\; \sin^2 2\theta_W {\rm
K}^2_{Z\gamma \gamma}
\nonumber\\
&\times&\left({\vec E}_{\theta}^2 + {\vec B}_{\theta}^2 + 42\left(
(E_\theta^3)^2+(B_\theta^3)^2\right)\right)\,. \label{eqn1p}
\end{eqnarray}
In order to estimate the scale of noncommutativity $\Lambda_{\rm NC}$ from
$\Gamma_{Z\rightarrow \gamma\gamma}$
we first have  to discuss the measurements.
The experimental lower bound for the partial width $\Gamma_{Z \rightarrow \gamma\gamma}$,
obtained from a previous experiment of the $e^+e^-\rightarrow \gamma\gamma$ annihilation,
is $< 5.2\times 10^{-5}$ GeV \cite{L3,DELPHI}, which gives the upper bound
$\Lambda_{\rm NC}>110\;\, \rm GeV$.

However, new experimental possibilities at LHC change the above bound,
lifting it up substantially.
According to the CMS Physics Technical Design Report \cite{CMS1},
around $10^7\;\, Z\to e^+ e^-$ events are expected to be recorded with $10 \;\,fb^{-1}$ of the data.
From this one can estimate the expected number of $Z \to \gamma\gamma$ events per $10 \;\,fb^{-1}$.
Assuming that $BR(Z \to \gamma\gamma) \sim 10^{-8}$ and using $BR(Z \to e^+e^-) = 3 \times 10^{-2}$,
we may expect to have $\sim 3$ events of $Z \to \gamma\gamma$ with $10\; fb^{-1}$.
Now the question is: What would be the background from $Z \to e^+e^-$ when the electron radiates
a very high-energy Bremsstrahlung photon in the beam pipe or in the first layer(s) of the Pixel
Detector and is thus lost for the tracker reconstruction?
In that case electron would not be reconstructed and it would be misidentified as a
photon. The probability of such an event should be evaluated from the
full detector simulation. According to the CMS note \cite{CMS2} which studies the $Z \to e^+e^-$
background for $Higgs \to \gamma\gamma$, the probability
to misidentify the electron as a photon is huge (see Fig. 3 in \cite{CMS2}) but
the situation can be improved applying more stringent selections to the photon candidate when
searching for $Z \to \gamma\gamma$ events \cite{Nikitenko}. However, the  irreducible
di-photon background (Fig. 3 in \cite{CMS2}) might also kill the signal.
In that case, one can only set the upper limits to the scale of noncommutativity from
the $Z \to \gamma\gamma$ branching ratio.

In accord with the analysis of the LHC experimental expectations \cite{CMS1,CMS2,Nikitenko} it
is bona fide reasonable to assume that the lower bound for the branching ratio is
$BR(Z \rightarrow \gamma\gamma) \stackrel{<}{\sim} 10^{-8}$.
Next, choosing the lower central value of $|K_{Z\gamma\gamma}|=0.05$, from
the figures and the Table in \cite{Duplancic:2003hg}, we find that
the upper bound to the scale of noncommutativity is
$\Lambda_{\rm NC}\stackrel{>}{\sim}1.0\;\,\rm TeV$
for ${\vec E}_{\theta}^2 + {\vec B}_{\theta}^2\simeq 1$.

Clearly, the measurement of the $Z \rightarrow \gamma\gamma$ decay branching ratio
would fix the quantity $|K_{Z\gamma\gamma}/\Lambda_{\rm NC}^2|$, while
the inclusion of other triple gauge boson
interactions through $2\to 2$ scattering experiments \cite{Ohl:2004tn} would sufficiently
reduce the available parameter space of our model by determining
more precisely the relations among the couplings
${\rm K}_{\gamma\gamma\gamma}$, ${\rm K}_{Z \gamma\gamma}$,
${\rm K}_{ZZ\gamma}$, ${\rm K}_{ZZZ}$, ${\rm K}_{WW\gamma}$, and ${\rm K}_{WWZ}$.

Let us summarize our results and compare to those obtained previously.

(A) The first calculation \cite{Mocioiu:2000ip} was performed
within a different model which  has different symmetries
in comparison with ours. In particular,  the model does not possess the commutative gauge invariance.
Note also that the  $Z\to\gamma\gamma$ rate obtained in \cite{Mocioiu:2000ip}
\begin{equation}
\Gamma_{Z\to\gamma\gamma} =\frac{\alpha}{144}\cos ^4\theta_W M_Z^5\sum_i\theta_{0i}^2 ,
\end{equation}
imposing the  unitarity of the theory in the usual manner,
$\theta^{0i} = 0$, \cite{Seiberg:2000gc,Gomis:2000zz}, vanishes.

(B) The partial width for the same process was obtained in \cite{Behr:2002wx,Duplancic:2003hg}
in the framework of similar theories, which, however, were not renormalizable. The present
results for the
 partial widths $\Gamma_{Z\rightarrow \gamma\gamma}$ and
$\Gamma^3_{Z\rightarrow \gamma\gamma}$ are about three times
larger than that in \cite{Behr:2002wx,Duplancic:2003hg} and consistently symmetric with respect
to time-space  and space-space noncommutativity.
In the polarized rate (\ref{eqn1p}) the third components
($(\theta^{03})^2+(\theta^{12})^2$)
are enhanced relative to the other two components by a large factor, as expected.
Also, the rate (\ref{eqn1p}) is enhanced
by a factor of $\sim$ 3 with respect to the total rate (\ref{eqn3}).

(C) The upper limit to the scale of noncommutativity $\Lambda_{\rm NC}\stackrel{>}{\sim} 1$
TeV is significantly higher than in \cite{Behr:2002wx}.
This bound is now firmer owing
to the regular behavior of the triple gauge boson
interactions (\ref{L1}-\ref{L4})
with respect to the one-loop renormalizability  of the NCSM
gauge sector.

(D) Finally, let us add that after 10 years of LHC running the integrated luminosity is expected
to reach $\sim 1000 \;\,fb^{-1}$,  \cite{CMS2}.
This means that for the assumed $BR(Z \to \gamma\gamma) \sim 10^{-8}$ we should have $\sim 300$
events of $Z \to \gamma\gamma$, that is we should be well above the background. On the other hand,
this result can also be understood as $\sim 3$ events with the $BR(Z \to \gamma\gamma) \sim 10^{-10}$,
which lifts the scale of noncommutativity up by a factor of $\sim 3$.

Therefore, with a more stringent selection of photon candidates and
if the irreducible di-photon contamination becomes controllable,
the $Z \rightarrow \gamma\gamma$ decay will become a clean signature of
space-time noncommutativity at very short distances in LHC experiments.

{\bf Acknowledgments}\\
We would like to thank A. Nikitenko for helpful discussions.
This work was supported by the Croatian Ministry of Science, Education and
Sport project 098-0982930-2900
and by the Serbian Ministry of Science project 141036.


\begin{thebibliography}{99}
%
%
\bibitem{Buric:2006wm}
  M.~Buric, V.~Radovanovic and J.~Trampetic,
  JHEP {\bf 0703}, 030 (2007)
  [arXiv:hep-th/0609073].
  %
\bibitem{Seiberg:1999vs}
N. Seiberg and E. Witten,
JHEP {\bf 09} (1999) 032 [hep-th/9908142].
%
\bibitem{Madore:2000en}
J. Madore, S. Schraml, P.~Schupp and J.~Wess,
Eur. Phys. J. C{\bf 16} (2000) 161 [hep-th/0001203];
  %
\bibitem{Maja}
M.~Buric and V.~Radovanovic,
  JHEP {\bf 0210} (2002) 074
  [arXiv:hep-th/0208204];
ibid
  JHEP {\bf 0402} (2004) 040
  [arXiv:hep-th/0401103];
%
  Class.\ Quant.\ Grav.\  {\bf 22} (2005) 525
  [arXiv:hep-th/0410085].
%
%
\bibitem{Bichl:2001cq}
A.~Bichl, J. Grimstrup, H. Grosse, L. Popp, M. Schweda and R. Wulkenhaar,
JHEP {\bf 06} (2001) 013 [hep-th/0104097];
%
J.~M.~Grimstrup and R.~Wulkenhaar,
Eur.\ Phys.\ J.\ C {\bf 26} (2002) 139
  [arXiv:hep-th/0205153];
 R.~Wulkenhaar,
 JHEP {\bf 0203} (2002) 024
 [arXiv:hep-th/0112248].
  %
\bibitem{Buric:2005xe}
  M.~Buric, D.~Latas and V.~Radovanovic,
  JHEP {\bf 0602} (2006) 046
  [arXiv:hep-th/0510133].
    %
\bibitem{Latas:2007eu}
  D.~Latas, V.~Radovanovic and J.~Trampetic,
  arXiv:hep-th/0703018.
%
%
%
\bibitem{Brandt:2003fx}
  C.~P.~Martin,
  Nucl.\ Phys.\ B {\bf 652}, 72 (2003)
  [arXiv:hep-th/0211164];
F. Brandt, C.P. Martin and F. Ruiz Ruiz,
JHEP {\bf 07} (2003) 068
[hep-th/0307292].
%
\bibitem{Yang} C. N. Yang, Phys. Rev. {\bf 77}, 242 (1950).
%
\bibitem{Hewett:2000zp}
J.~L.~Hewett, F.~J.~Petriello and T.~G.~Rizzo,
Phys.\ Rev.\ D {\bf 64} (2001) 075012
[hep-ph/0010354];
%
\bibitem{Ohl:2004tn}
T.~Ohl and J.~Reuter,
Phys. Rev. D{\bf 70} (2004) 076007 [hep-ph/0406098];
  A.~Alboteanu, T.~Ohl and R.~Ruckl,
  PoS {\bf HEP2005} (2006) 322
  [arXiv:hep-ph/0511188];
  A.~Alboteanu, T.~Ohl and R.~Ruckl,
  Phys.\ Rev.\ D {\bf 74}, 096004 (2006)
  [arXiv:hep-ph/0608155].
%
\bibitem{Abbiendi:2003wv}
  G.~Abbiendi {\it et al.}  [OPAL Collaboration],
  Phys.\ Lett.\ B {\bf 568}, 181 (2003)
  [hep-ex/0303035].
 %
\bibitem{Peter}
P.~Schupp, J.~Trampeti\'{c}, J.~Wess and G.~Raffelt,
Eur.\ Phys.\ J.\ C {\bf 36} (2004) 405
[hep-ph/0212292].
%
P.~Minkowski, P.~Schupp and J.~Trampeti\'{c},
Eur.\ Phys.\ J.\ C {\bf 37} (2004) 123
[hep-th/0302175].
\bibitem{Carroll:2001ws}
  S.~M.~Carroll, J.~A.~Harvey, V.~A.~Kostelecky, C.~D.~Lane and T.~Okamoto,
  Phys.\ Rev.\ Lett.\  {\bf 87}, 141601 (2001)
  [arXiv:hep-th/0105082].
%
\bibitem{Mocioiu:2000ip}
  I.~Mocioiu, M.~Pospelov and R.~Roiban,
  Phys.\ Lett.\ B {\bf 489}, 390 (2000)
  [arXiv:hep-ph/0005191];
\bibitem{Josip}
J. Trampeti\'c,
Acta Phys. Polon. B{\bf 33}
(2002) 4317 [hep-ph/0212309];
ibid
arXiv:0704.0559 [hep-ph];
%
  B.~Melic, K.~Passek-Kumericki and J.~Trampetic,
  Phys.\ Rev.\ D {\bf 72} (2005) 054004
  [arXiv:hep-ph/0503133]; ibid
  Phys.\ Rev.\ D {\bf 72} (2005) 057502
  [arXiv:hep-ph/0507231];
  %
\bibitem{Calmet:2001na}
X.~Calmet, B.~Jur\v{c}o, P.~Schupp, J.~Wess and M.~Wohlgenannt,
Eur.~Phys.~J. C{\bf 23} (2002) 363
[hep-ph/0111115].
%
\bibitem{Aschieri:2002mc}
P. Aschieri, B.~Jur\v{c}o, P.~Schupp and J.~Wess,
Nucl. Phys. B{\bf 651} (2003) 45 [hep-th/0205214].
%
\bibitem{Behr:2002wx}
W.~Behr, N.G. Deshpande, G. ~Duplan\v{c}i\'{c}, P.~Schupp, J.~Trampeti\'{c} and
J.~Wess,
Eur. Phys. J. C{\bf 29} (2003) 441 [hep-ph/0202121];
%
\bibitem{Duplancic:2003hg}
G.~Duplan\v{c}i\'{c}, P.~Schupp and J.~Trampeti\'{c},
Eur.~Phys. J. C{\bf 32} (2003) 141 [hep-ph/0309138].
%
\bibitem{Melic:2005fm}
  B.~Melic, K.~Passek-Kumericki, J.~Trampetic, P.~Schupp and M.~Wohlgenannt,
  Eur.\ Phys.\ J.\ C {\bf 42} (2005) 483
  [arXiv:hep-ph/0502249]; ibid
  Eur.\ Phys.\ J.\ C {\bf 42} (2005) 499
  [arXiv:hep-ph/0503064].
\bibitem{Martin:2006gw}
  C.~P.~Martin, D.~Sanchez-Ruiz and C.~Tamarit,
  JHEP {\bf 0702}, 065 (2007)
  [arXiv:hep-th/0612188].
  %
%


%
%
%
  \bibitem{L3}  L3 Collaboration, M. Acciari et al., Phys. Lett. {\bf B436}, 187 (1999);
              L3 Collaboration, M. Acciari et al., Phys. Lett. {\bf B489}, 55 (2000);
              L3 Note 2672 (July 2001).

\bibitem{DELPHI} DELPHI Collaboration,
                 DELPHI 2001-097, CONF 525.

%
\bibitem{CMS1} CMS Physics Technical Design Report, Vol.1.  CERN/LHCC 2006-001.
%
\bibitem{CMS2} M. Pieri et al., CMS Note 2006/112.
%
\bibitem{Nikitenko}  A. Nikitenko, private communications.
%
\bibitem{Seiberg:2000gc}
  N.~Seiberg, L.~Susskind and N.~Toumbas,
  JHEP {\bf 0006}, 044 (2000)
  [arXiv:hep-th/0005015].
\bibitem{Gomis:2000zz}
  J.~Gomis and T.~Mehen,
  Nucl.\ Phys.\ B {\bf 591}, 265 (2000)
  [arXiv:hep-th/0005129].
%
\end{thebibliography}
\end{document}